# OFFLINE SOFTWARE AND COMPUTING FOR THE SPD EXPERIMENT

**V. Andreev[1] [0000-0002-5492-6920], A. Belova[2] [], A. Galoyan[2] [], S. Gerassimov[1] [0000-0001-7780-8735], G. Golovanov[2] [], P. Goncharov[2] [0000-0003-4199-6329], A. Gribowsky[2] [], D. Maletic[3] [0000-0002-9163-6703], A. Maltsev[2] [], A. Nikolskaya[4] [0000-0001-9113-8635], D. Oleynik[2] [0000-0002-8862-770X], G. Ososkov[2] [0000-0002-3223-6953], A. Petrosyan[2] [0000-0003-2166-7894], E. Rezvaya[2] [], E. Shchavelev[4] [0000-0002-7016-6028], A.Tkachenko[2] [], V. Uzhinsky[2] [0000-0002-1634-3468], A. Verkheev[2] [0000-0002-0578-1323], A. Zhemchugov[2] a [0000-0002-3360-4965]**

[1] *P.N. Lebedev Physical Institute of the Russian Academy of Sciences, 53 Leninskiy Prospekt,119991, Moscow, Russia*

[2] *Joint Institute for Nuclear Research, 6 Joliot-Curie, Dubna, Moscow region, 141980 Russia*

[3] *Institute of Physics Belgrade, Pregrevica 118, Belgrade, Serbia*

[3] *Saint Petersburg State University, 7-9 Universitetskaya emb., Saint Petersburg, 199034, Russia*

E-mail: [a] zhemchugov@jinr.ru

The SPD (Spin Physics Detector) is a planned spin physics experiment in the second interaction point of the NICA collider that is under construction at JINR. The main goal of the experiment is the test of basic of the QCD via the study of the polarized structure of the nucleon and spin-related phenomena in the collision of longitudinally and transversely polarized protons and deuterons at the center-of-mass energy up to 27 GeV and luminosity up to $10^{32}$ 1/(cm$^2$ s). The data rate at the maximum design luminosity is expected to reach 0.2 Tbit/s. Current approaches to SPD computing and offline software will be presented. The plan of the computing and software R&D in the scope of the SPD TDR preparation will be discussed.

Vladimir Andreev, Anna Belova, Aida Galoyan, Sergey Gerassimov, Georgy Golovanov, Pavel Goncharov, Alexander Gribowsky, Dimitrije Maletic, Andrey Maltsev, Anastasiya Nikolskaya, Danila Oleynik, Gennady Ososkov, Artem Petrosyan, Ekaterina Rezvaya, Egor Shchavelev, Artur Tkachenko, Vladimir Uzhinsky, Alexander Verkheev, Alexey Zhemchugov

# 1. Introduction

The SPD (Spin Physics Detector) is a planned spin physics experiment in the second interaction point of the NICA collider that is under construction at JINR. The main goal of the experiment is the test of basic of the QCD via the study of polarized structure of nucleon and spin-related phenomena in the collision of longitudinally and transversely polarized protons and deuterons at the center-of-mass energy up to 27 GeV and luminosity up to $10^{32}$ cm$^{-2}$ s$^{-1}$. Transverse Momentum-Dependent partonic distributions (TMD PDFs) for gluons is the main goal of the experiment. They will be accessed in such hard processes as the production of the open charm, charmonia and prompt photons. The SPD detector is planned as a universal 4π spectrometer based on modern technologies. It will include such subsystems as a silicon vertex detector, a straw tube-based main tracker, a time-of-flight system, an electromagnetic calorimeter and a muon identification system. A total number of the detector channels in SPD is about 500 000, with the major part coming from the vertex detector. Assuming that all sub-detectors are in operation, the raw data flow was estimated as 10–20 GB/s. This poses a significant challenge both to the DAQ system and to the offline computing system and data processing software. Taking into account recent advances in the computing hardware and software, the investment in the research and development necessary to deploy software to acquire, manage, process, and analyze the data recorded is required along with the physics program elaboration and the detector design.

# 2. SPD computing model

Expected event rate of the SPD experiment is about 3 MHz (*pp* collisions at √s = 27 GeV and $10^{32}$ cm$^{-2}$s$^{-1}$ design luminosity). This is equivalent to the raw data rate of 20 GB/s or 200 PB/year, assuming the detector duty cycle is 0.3, while the signal-to-background ratio is expected to be in order of $10^{-5}$. Taking into account the bunch crossing rate of 12.5 MHz, one may conclude that pile-up probability cannot be neglected.

The key challenge of the SPD Computing Model is the fact, that no simple selection of physics events is possible at the hardware level, because the trigger decision would depend on measurement of momentum and vertex position, which requires tracking. Moreover, the free-running DAQ provides a continuous data stream, which requires a sophisticated unscrambling prior building individual events. That is the reason why any reliable hardware-based trigger system turns out to be over-complicated and the computing system will have to cope with the full amount of data supplied by the DAQ system. This makes a medium-scale setup of SPD a large-scale data factory.

The continuous data reduction is a key point in the SPD computing. While simple operations like noise removal can be done yet by DAQ, it is an online filter that is aimed at fast partial reconstruction of events and data selection, thus being a kind of a software trigger. The goal of the online filter is to decrease the data rate at least by a factor of 50 so that the annual upgrowth of data including the simulated samples stays within 10 PB. Then, data are transferred to the Tier-1 facility, where full reconstruction takes place and the data is stored permanently. Two reconstruction cycles are foreseen. The first cycle includes reconstruction of some fraction of each run necessary to study the detector performance and derive calibration constants, followed by the second cycle of reconstruction of full data sample for physics analysis. The data analysis and Monte-Carlo simulation will likely run at the remote computing centers (Tier-2s). Given the large data volume, a thorough optimization of the event model and performance of reconstruction and simulation algorithms are necessary.

Taking into account recent advances in the computing hardware and software, the investment in the research and development necessary to deploy software to acquire, manage, process, and analyze the data recorded is required along with the physics program elaboration and the detector design. While the core elements of the SPD computing system and offline software now exist as prototypes, the system as a whole with capabilities such as described above is in the conceptual design stage and information will be added as it is developed.

## 3. Online filter

The SPD online filter facility will be a high-throughput system which will include heterogeneous computing platforms similar to many high-performance computing clusters. The computing nodes will be equipped with hardware acceleration. The software framework will provide the necessary abstraction so that common code can deliver the selected functionality on different platforms.

The main goal of the online filter is a fast reconstruction of the SPD events and suppression of the background ones at least by a factor of 50. This requires fast tracking and fast clustering in the electromagnetic calorimeter, followed by reconstruction of event from a sequence of time slices and an event selection (software trigger). Several consecutive time slices shall be considered, tracker data unpacked and given for a fast tracking. The result of the fast track reconstruction is the number of tracks, an estimate of their momentum and an estimate of primary vertex (to distinguish between tracks belonging to different collisions). Using this outcome, the online filter should combine information from the time slices into events and add a trigger mark. The events shall be separated in several data streams using the trigger mark and an individual prescale factor for each stream is applied.

One of the most important aspects of this chain is the recognition of particle tracks. Traditional tracking algorithms, such as the combinatorial Kalman filter, are inherently sequential, which makes them rather slow and hard to parallelize on modern high-performance architectures (graphics processors). As a result, they do not scale well with the expected increase in the detector occupancy during the SPD data taking. This is especially important for the online event filter, which should be able to cope with the extremely high data rates and to fulfill the significant data reduction based on partial event reconstruction "on the fly". The parallel resources like multicore CPU and GPU farms will likely be used as a computing platform, which requires the algorithms, capable of the effective parallelization, to be developed, as well as the overall cluster simulation and optimization.

Machine learning algorithms are well suited for multi-track recognition problems because of their ability to reveal effective representations of multidimensional data through learning and to model complex dynamics through computationally regular transformations, that scale linearly with the size of input data and are easily distributed across computing nodes. Moreover, these algorithms are based on the linear algebra operations and can be parallelized well using standard ML packages. This approach has already been applied successfully to recognize tracks in the BM@N experiment at JINR and in the BESIII experiment at IHEP CAS in China [1, 2]. In the course of the project an algorithm, based on recurrent neural networks of deep learning, will be developed to search for and reconstruct tracks of elementary particles in SPD data from the silicon vertex detector and the straw tube-based main tracker. The same approach will be applied to the clustering in the SPD electromagnetic calorimeter, and fast $\pi^0$ reconstruction. The caution is necessary, though, to avoid possible bias due to an inadequacy of the training data to the real ones, including possible machine background and the detector noise. A dedicated workflow that includes continuous learning and re-learning of neuron network, deployment of new versions of network and the continuous monitoring of the performance of the neural networks used in the online filter is necessary and needs to be elaborated.

Besides the high-level event filtering and corresponding data reduction, the online filter will provide input for the run monitoring by the shift team and the data quality assessment, as well as local polarimetry.

## 4. Computing system

The projected rate and amount of data produced by SPD prescribe to use high throughput computing solutions for the processing of collected data. It is the experience of a decade of the LHC computing that already developed a set of technologies mature enough for the building of distributed high-throughput computing systems for the experiments in high energy physics.

The 'online' part of computing systems for the SPD experiment, namely the online filter described above, is an integral part of experimental facilities, connected with the 'offline' part using a high throughput backbone network. The entry point to 'offline' facilities is a high capacity storage system, connected with 'online facility' through a multilink high-speed network. Data from high capacity storage at the Meshcheryakov Laboratory of Information Technologies will be copied to the tape-based mass storage system for long-term storage. At the same time, data from high capacity storage will be processed on different computing facilities as at JINR as in other collaborative institutions.

The hierarchy of offline processing facilities can be introduced:

- Tier 1 level facilities should provide high capacity long-term storage which will have enough capacity to store a full copy of primary data and a significant amount of important derived data;
- Tier 2 level facility should provide (transient) storage with capacity that will be enough for storing of data associated with a period of data taking;
- optional Tier 3 level are opportunistic resources, that can be used to cope with a pile-up of processing during some period of time or for special analysis.

Offline data processing resources are heterogeneous as on hardware architecture level so by technologies and at JINR site it includes batch processing computing farms, high-performance (supercomputer) facilities, and cloud resources. A set of middleware services will be required to have unified access to different resources.

Computing systems for NICA at JINR are naturally distributed. Experimental facilities and main data processing facilities placed across two JINR sites and, inter alia, managed by different teams. That causes some heterogeneity not only on hardware systems but also on the level of basic software: different OSs, different batch systems etc. Taking into account the distributed nature and heterogeneity of the existing infrastructure, and expected data volumes, the experimental data processing system must be based on a set of low-level services that have proven their reliability and performance. It is necessary to develop a high-level orchestrating system that will manage the low-level services. The main task of that system will be to provide efficient, highly automated multi-step data processing following the experimental data processing chain.

The Unified Resource Management System is a IT ecosystem composed from the set of subsystem and services which should:

- unify the access to the data and compute resources in a heterogeneous distributed environment;
- automate most of the operations related to massive data processing;

- avoid duplication of basic functionality, through sharing of systems across different users (if it possible);
- as a result - reduce operational cost, increase the efficiency of usage of resources;
- transparent accounting of usage of resources.

Many distributed computing tools have already been developed for the LHC experiments and can be re-used in SPD. For the task management one can use PANDA [3] or DIRAC [4] frameworks. For the distributed data management RUCIO [5] package has been developed. For the massive data transfer FTS [6] can be used. Evaluation of these tools for the SPD experiment and their implementation within the SPD Unified Resource Management System is planned in scope of the TDR preparation.

## 5. Offline software

Offline software is a toolkit for event reconstruction, Monte-Carlo simulation and data analysis. Linux is chosen as a base operating system.

Currently, the offline software of the SPD experiment – SpdRoot – is derived from the FairRoot software [7] and it is capable of Monte Carlo simulation, event reconstruction, data analysis and visualization. The SPD detector description is flexible and based on the ROOT geometry package. Proton-proton collisions are simulated using a multipurpose generator Pythia8 [8]. Deuteron-deuteron collisions are simulated using a modern implementation of the FRITIOF model [9, 10], while UrQMD [11, 12] generator is used to simulate nucleus-nucleus interactions. Transportation of secondary particles through the material and the magnetic field of the SPD setup and the simulation of detector response is provided by Geant4 toolkit [13, 14, 15]. Track fitting is carried out on the base of GenFit toolkit [16, 17] while the KFparticle package [18] is used to reconstruct secondary vertices. The central database is going to be established to keep and distribute run information, slow control data and calibration constants.

Recent developments in computing hardware resulted in the rapid increase in potential processing capacity from increases in the core count of CPUs and wide CPU registers. Alternative processing architectures have become more commonplace. These range from the multi-core architecture based on x86_64 compatible cores to numerous alternatives such as other CPU architectures (ARM, PowerPC) and special co-processors/accelerators: (GPUs, FPGA, etc). For GPUs, for instance, the processing model is very different, allowing a much greater fraction of the die to be dedicated to arithmetic calculations, but at a price in programming difficulty and memory handling for the developer that tends to be specific to each processor generation. Further developments may even see the use of FPGAs for more general-purpose tasks.

The effective use of these computing resources may provide a significant improvement in offline data processing. However, the offline software should be capable to do it by taking advantage of concurrent programming techniques, such as vectorization and thread-based programming. Currently, the SPD software framework, SpdRoot, cannot use these techniques effectively. The studies of the concurrent-capable software frameworks (e.g. ALFA [19], Key4Hep [20]) are needed to provide input for the proper choice of the offline software for Day-1 of the SPD detector operation, as well as a dedicated R&D effort to find proper solutions for the development of efficient cross-platform code.

A git-based infrastructure for the SPD software development already established at JINR [21].

## 4. Conclusion

The expected SPD data rate of 0.2 Tbit/s at the maximum design luminosity poses a significant challenge to the DAQ system, to the online event filter, and to the offline computing system and data processing software. Fast event reconstruction based on deep learning algorithms, distributed computing, and extensive use of concurrent programming techniques, such as vectorization and thread-based programming, are the key points in the SPD data processing paradigm. Taking into account recent advances in computing hardware and software, the investment in the research and development necessary to deploy software to acquire, manage, process, and analyze the data recorded is required along with the physics program elaboration and the detector design.

## References

[1] D. Baranov, S. Mitsyn, P. Goncharov, G. Ososkov, The Particle Track Reconstruction based on deep neural networks // EPJ Web Conf. 214 (2019) 06018

[2] G. Ososkov, et al., Tracking on the BESIII CGEM inner detector using deep learning // Computer Research and Modeling 10 (20) 1–24

[3] F. B. Megino, et al., PanDA: Evolution and Recent Trends in LHC Computing // Procedia Comput. Sci. 66 (2015) 439–447

[4] F. Stagni, A. Tsaregorodtsev, L. Arrabito, A. Sailer, T. Hara, X. Zhang, DIRAC in Large Particle Physics Experiments // J. Phys. Conf. Ser. 898 (9) (2017) 092020

[5] M. Barisits, T. Beermann, F. Berghaus, et al., Rucio: Scientific data management // Comput. Softw. Big Sci. 3 (2019) 11

[6] A. Frohner, J.-P. Baud, R. M. Garcia Rioja, G. Grosdidier, R. Mollon, D. Smith, P. Tedesco, Data management in EGEE // J. Phys. Conf. Ser. 219 (2010) 062012

[7] M. Al-Turany, D. Bertini, R. Karabowicz, D. Kresan, P. Malzacher, T. Stockmanns, F. Uhlig, The FairRoot framework // J. Phys. Conf. Ser. 396 (2012) 022001

[8] T. Sjöstrand, S. Ask, J. R. Christiansen, R. Corke, N. Desai, P. Ilten, S. Mrenna, S. Prestel, C. O. Rasmussen, P. Z. Skands, An introduction to PYTHIA 8.2 // Comput. Phys. Commun. 191 (2015) 159–177

[9] B. Andersson, G. Gustafson, B. Nilsson-Almqvist, A Model for Low p(t) Hadronic Reactions, with Generalizations to Hadron - Nucleus and Nucleus-Nucleus Collisions // Nucl. Phys. B 281 (1987) 289–309

[10] B. Nilsson-Almqvist, E. Stenlund, Interactions Between Hadrons and Nuclei: The Lund Monte Carlo, Fritiof Version 1.6 // Comput. Phys. Commun. 43 (1987) 387

[11] S. Bass, et al., Microscopic models for ultrarelativistic heavy ion collisions // Prog. Part. Nucl. Phys. 41 (1998) 255–369.

[12] M. Bleicher, et al., Relativistic hadron hadron collisions in the ultrarelativistic quantum molecular dynamics model // J. Phys. G 25 (1999) 1859–1896

[13] S. Agostinelli, et al., GEANT4–a simulation toolkit // Nucl. Instrum. Meth. A 506 (2003) 250–303

[14] J. Allison, et al., Geant4 developments and applications // IEEE Trans. Nucl. Sci. 53 (2006) 270

[15] J. Allison, et al., Recent developments in Geant4 // Nucl. Instrum. Meth. A 835 (2016) 186–225

[16] J. Rauch, T. Schlüter, GENFIT — a Generic Track-Fitting Toolkit // J. Phys. Conf. Ser. 608 (1) (2015) 012042

[17] https://github.com/GenFit/GenFit

[18] S. Gorbunov, I. Kisel, Reconstruction of decayed particles based on the Kalman filter // Tech. Rep. CBM-SOFT-note-2007-003, CBM Collaboration (2007)

[19] M. Al-Turany, et al., ALFA: The new ALICE-FAIR software framework, J. Phys. Conf. Ser. 664 (7) (2015) 072001

[20] Key4hep software // https://key4hep.github.io/key4hep-doc/index.html

[21] https://git.jinr.ru/nica/spdroot